\documentclass[twocolumn,aps,prl,floats,showpacs]{revtex4}
\newcommand{\beq}{\begin{eqnarray}}
\newcommand{\eeq}{\end{eqnarray}}

\usepackage{epsfig}
\usepackage{dcolumn}
\usepackage{bm}
\usepackage{amsfonts}
\usepackage{latexsym}

\begin{document}
\bibliographystyle{prsty} \title{Electron Transport
  in Granular Metals} \draft
\author{Alexander Altland$^{1}$, Leonid I. Glazman$^{2,3}$, and Alex  Kamenev$^{2}$}

\address{$^1$Institut f\"ur
  theoretische Physik, Z\"ulpicher Str. 77, 50937 K\"oln, Germany}
\address{$^2$Department of Physics, and $^3$W. I. Fine Theoretical
  Physics Institute, University of Minnesota, Minneapolis, MN 55454,
  USA}

\date{\today}
\begin{abstract}
  {\rm We consider thermodynamic and transport properties of a long
    granular array with strongly connected grains (inter-grain
    conductance $g\gg 1$.)  We find that the system exhibits activated
    behavior of conductance and thermodynamic density of states $\sim
    \exp\{-T^\ast/T\}$ where the gap, $T^\ast$, is parametrically
    larger than the energy at which conventional perturbation
    theory breaks down.  The scale $T^\ast$ represents
    energy needed to create a long single--electron charge soliton
    propagating through the array.}
\end{abstract}
 \maketitle
\bigskip

The low-temperature conductivity of granular materials continues
to attract attention, from both experimentalists~\cite{Gerber97}
and theorists~\cite{Golubev92,Beloborodov01}. From a conceptual
point of view, an attractive feature of these systems is the
possibility to separately control the effects of electron
interaction and quantum interference. A particularly interesting
situation is realized in arrays with large inter-granular
conductance, $g\gg 1$ (in units of $e^2/h$), and large grain size
(small electron mean level spacing, $\delta$, in the grains). In
the limit $g\delta \ll T$ electron transport in such systems
becomes purely inelastic and long range quantum coherence is
inhibited~\cite{Beloborodov01}. As we show below, under these
conditions interaction effects {\it alone} lead to an exponential
suppression of conductivity, which is fully amenable to analytical
treatment.

At high temperatures, the conductivity  of a granular array is
Ohmic, $\sigma=g$ (hereinafter the length of the system is
measured in the number of grains). At lower temperatures
Altshuler-Aronov interaction  corrections~\cite{Altshuler85} begin
to impede the conduction behavior. For ``inelastic'' arrays this
correction was found~\cite{Golubev92} to be $\delta\sigma=-\ln
E_c/T$, where $E_c$ is the charging energy of the individual
grains. Comparison with the Ohmic contribution shows that this
{\em perturbative} correction is small as long as $T>\tilde
E_c\equiv E_c e^{-g}$. At the same energy scale, $\tilde E_c$, a
{\em single} grain connected to external leads would cross over to
the strong Coulomb blockade regime
\cite{Zaikin91,Grabert96,Nazarov99}.

In this paper we show that the conductivity of a 1$d$ {\em array} of
grains crosses over to a manifestly insulating (activated)
behavior at a parametrically {\em larger} temperature, $T^\ast\gg
\tilde E_c$. Below the crossover, the conductivity is
exponentially small:
\begin{equation}
\label{result}
 \sigma = g\, \exp\left(-{T^\ast\over T}\right)\,, \hskip 1cm  T\lesssim T^\ast\,,
\end{equation}
as characteristic for insulators. The size of the gap, $T^\ast$,
is model-dependent.  For arrays with vanishing background charge,
$q$, at each grain, we find $T^\ast \sim E_c e^{-g/4}$, while in
the case of random background charges $T^\ast \sim E_c e^{-g/2}$.
In either case $T^\ast \gg \tilde E_c$. In the case of $q=0$, the
thermodynamic density of states (DOS) is suppressed along with the
conductivity. Note that Eq.~(\ref{result}) is {\em not} a result
of phonon--mediated hopping, but is a consequence of interactions
between electrons only.

The reason why the scale $T^\ast$ and Eq.~(\ref{result}) were
overlooked previously is that they are not visible in standard
perturbative expansions in $1/g\ll 1$. In the conventional
formulation of the theory in terms of {\em voltage}
fluctuations~\cite{Schon90}, Eq.  (\ref{result})  comes from
including large, topologically non--trivial fluctuations
(instantons). Proliferation of instantons leads to insulating
behavior at temperatures $T^\ast \gg \tilde E_c$, where Gaussian
fluctuations are still small. Notice that for a {\it single} grain
instantons affect the conductance only at much lower temperatures
$T\approx \tilde E_c$~\cite{Golubev96}. However, contrary to a
single dot, an extended array provides a large 'entropic volume'
for the formation of instantons, which substantially  increases
the characteristic temperature. We shall return to a quantitative
discussion of this picture below.

It turns out, however, that the effect is more naturally explained
using a language of {\em charge} fluctuations. It is known that
even a highly conducting barrier retains some ability to pin the
charge on a single grain~\cite{Zaikin91,Grabert96,Nazarov99}. This
mechanism is drastically enhanced in the array geometry, where it
bears similarity to the pinning of charge density waves by a
periodic potential. The elementary mobile excitations in this
system are extended solitons of unit charge.  Their activation
energy, $T^\ast$, is given by the geometric mean of the pinning
strength and inverse charge compressibility (grain capacitance).
Our main result, Eq. (\ref{result}), simply reflects the thermal
density of such charge solitons.

To quantify this latter picture we consider a generalization of a
model previously employed to study quantum dots~\cite{Flensberg93}.
Its simplest version treats the grains coupled by a single conducting
channel and therefore has $g\lesssim 1$.  (We shall show later that
the predictions derived from it survive generalization to the
complementary case $g\gg 1$.)  The model is formulated in terms of a
charge displacement field, $\theta_j(\tau)$, where
$\theta_{j+1}-\theta_j = N_j$ is the charge on the $j$-th grain.  In
the absence of  backscattering at the contacts, the action reads
\begin{equation}
 \label{S0}
 S_0=\sum\limits\limits_{j=1}^{M-1} {1\over T}\sum\limits_m
 \left[E_c\,(\theta_{j+1} -\theta_j-q)^2 +\pi |\omega_m| \theta_j^2
 \right]\,,
\end{equation}
where the first term represents the charging energy of the grains,
while the second originates from integrating out the continuum of the
electronic degrees of freedom. Backscattering at the
inter--granular junctions is described \cite{Flensberg93} by a
nonlinear term to be added to the action (\ref{S0}): $S_{\rm bs} =
{Dr\over \pi} \sum_j\int d\tau \cos (2\pi \theta_j(\tau))$.  Here, $r$
is the reflection amplitude and $D\gg E_c$ the effective bandwidth of
the model.

A crucial observation that makes the problem solvable   is that
even for $r=0$ the {\em quantum}  fluctuations of $\theta_j(\tau)$
do {\em not}~\cite{foot1} diverge in the limit $T\to 0$:
\begin{equation}
\label{fluctuations}
 \langle \theta_j(\tau)^2 \rangle ={T\over M}\!\! \sum\limits_{k=0}^{M-1}\!\! 
\sum\limits_{m\neq 0} \frac{e^{-|\omega_m|/D} }{E_k
+\pi |\omega_m|} = {1\over 2\pi^2}\, \ln{\pi D\over e^{\bf C} E_c}
\, ,
\end{equation}
where $E_k =4E_c \sin^2 (\pi k/2M)$ is the excitation spectrum
defined by Eq.~(\ref{S0}), and ${\bf C}\approx 0.577$ is the Euler
constant. One can thus safely integrate out these fluctuations, to
arrive at a sine-Gordon type action that involves only the {\em
classical} (zero Matsubara) component of the field:
\begin{equation}
\label{eq:2}
S[\theta] ={E_c\over T} \sum\limits_{j=1}^{M-1} \left[ (\theta_{j+1}-\theta_j-q)^2 - 2\gamma \cos \, (2\pi
\theta_j) \right]\, ,
\end{equation}
where $\gamma \equiv |r|e^{\bf C}/(2\pi^2)$. In the
multi--channel case the coupling constant generalizes
\cite{Nazarov99,Aleiner02} to $\gamma \sim\prod |r_s|$, where
$r_s$ is the reflection coefficient of the $s$th channel.
\begin{figure}
\centerline{\epsfxsize=3in\epsfbox{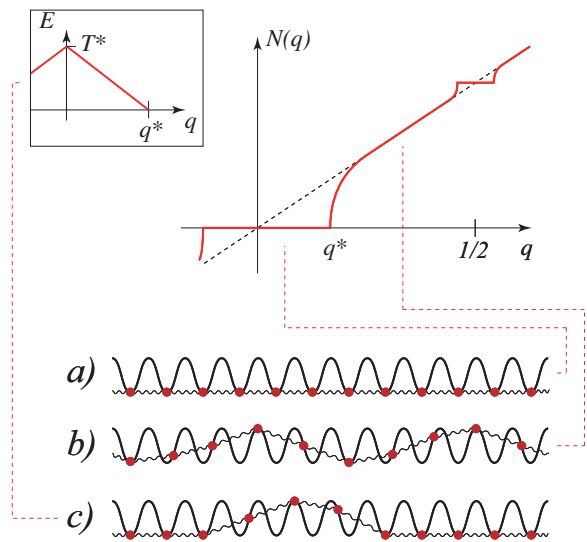}}
\vskip0.2in
\caption{Plot of the function (see text) $\bar N(q)$
  of an atomic chain in proximity to a periodic substrate. a) (b))
  commensurate (incommensurate) configuration, c) solitary
  excitation with its excitation energy (inset.)}
\label{fig:fk_model}
\end{figure}

Equation~(\ref{eq:2}) is known as the action of the
Frenkel--Kontorova model \cite{Chaikin}.  This model describes a
harmonic elastic chain of "atoms" with stiffness $E_c$, placed on
top of a periodic "substrate" potential with the amplitude
$2\gamma E_c$. The "incommensurability parameter" $q$ represents
the periodicity mismatch between the chain and the substrate.  For
small values of $q$ the system will find it favorable to retain a
commensurate state (cf.  Fig. \ref{fig:fk_model} a)), i.e. the
chain will stretch a little so as to still benefit from an optimal
coupling to the substrate.  Setting $\theta_j = 0$, one finds that
the energy per atom in this configuration is given by $F_c = E_c
(q^2-2\gamma)$. At $|q|>2\gamma$ this energy becomes positive, and
the state with $\theta_j=0$ can not persist as the lowest-energy
state (e.g. it is obviously  less favorable than the
incommensurate state with $\theta_j\approx jq $ and $F_i\approx
0$). Indeed, in the limit of weak periodic potential, the
transition between the commensurate, (Fig. \ref{fig:fk_model}, a)
and incommensurate (Fig. \ref{fig:fk_model}, b) phases occurs at
$|q|=q^\ast\equiv\sqrt{2\gamma}$. For the average number of
electrons per grain, $\bar N(q) \equiv q-\partial_q F/(2E_c)$, one
thus expects: $\bar N(q) =0$ for $|q|\leq q^\ast$ (insulator) and
$ \bar N \to q$ for $|q|>q^\ast$ (metal).

The relevant {\it thermal excitations} superimposed on the
commensurate ground state are so-called incommensurations (solitons in
the language of the sine-Gordon model) -- localized defects, where the
distortion $\theta_j$ 'climbs' over a maximum of the substrate
potential to relax back into a next minimum (cf.  Fig.
\ref{fig:fk_model} c)).  Minimizing the action (\ref{eq:2}), subject
to the boundary condition $\theta_{\pm\infty}=0(1)$, and using
the condition $\gamma \ll 1$, one finds that the action of one of these
solitons is given by $S_s=T^\ast(q)/T$, where
\begin{equation}
T^\ast(q)\stackrel{|q|<q^\ast}{=} 2\pi E_c(q^\ast -|q|)\, ; \hskip
1cm q^\ast=\sqrt{2\gamma}\, .
                                            \label{tstar}
\end{equation}
As a result, at $|q| < q^\ast$ the thermodynamic density of states
scales as $\partial_q \bar N(q) \sim \exp(-T^\ast(q)/T)$, i.e.
$T^\ast(q)$ is the excitation gap of the system. Consequently the
conductivity exhibits the same activation behavior, cf.
Eq.~(\ref{result}). Notice, that for $|q|=q^\ast$ the  gap
vanishes. In agreement with our earlier estimate, this signals a
proliferation of solitary excitations and the proximity of the
incommensurate phase.

A more thorough discussion of the system (cf. Ref.~\cite{Chaikin})
shows that insulating 'plateaus' along with superimposed solitary
excitations form not only around $q=0$, but also around  other
rational values of $q$. However, both the width of these plateaus
and the corresponding activation energies decrease for higher
rational fractions. Among the low lying rationals, $q=1/2$ plays a
particularly interesting role. Indeed, for a {\it single} grain,
$q= 1/2$ represents charge degeneracy point, where the system is
in a conducting state (Coulomb blockade peak).  Unexpectedly, the
{\it array} exhibits a very different behavior.  Using our current
language, $q=\pm 1/2$ is special in that the atoms of the
unperturbed chain alternatingly find themselves in minima/maxima
of the substrate potential. Under these conditions, energy can be
gained by building up an 'Peierls-distortion' of periodicity $2$
and modulation amplitude $\delta \theta_j\sim \gamma$. This
configuration is inert against small variations in $q$
(insulating) where, however, the width of the insulating plateau
estimates to only $\Delta q_{1/2} \sim \gamma$, i.e. it is much
smaller than $\Delta q_{0}\equiv 2q^\ast\sim \sqrt{\gamma}$.

The above discussion was based on the arguably artificial
assumption that the background charges in every grain are the
same. Under realistic conditions, though, one expects $q\to q_j$
to fluctuate. (The same applies to the tunneling conductances and
charging energies; we believe, however, that these latter
fluctuations are of lesser relevance.)  Let us briefly consider
the extreme limit where $q_j\in [0,1[$ on the different grains are
uniformly distributed statistically independent random variables.
For an undistorted chain, $\theta_{j+1}-\theta_j=q_j$, the
potential terms $2\gamma\cos 2\pi \theta_j$ vary randomly and the
energy per atom is zero on average. The system can then gain an
energy $\delta F\approx E_c\gamma^2$ per grain by slightly
distorting the chain, so that that
$\delta\theta_{j+1}-2\delta\theta_j +\delta\theta_{j-1} =
\gamma\sin(2\pi jq_j)$. In analogy with the "clean" case, the
excitation energy of this deformed state is expected to be $T^\ast
\sim \sqrt{E_c\delta F} \approx E_c\gamma$. (Note that this
mechanism closely related to the collective pinning of Abrikosov
lattices in type II superconductors \cite{Larkin}.)

Having discussed the charge pinning mechanism in the context of
the few channel model, we next turn to the generalization to
highly conducting arrays ($g\gg 1$). To this end we employ the
so-called Ambegoakar-Eckern-Sch\"on (AES) model \cite{Schon90}.
This  formalism  describes the system in terms of the quantum
phase, $\phi_j(\tau)$, conjugated to the charge
$\theta_{j+1}(\tau)-\theta_j(\tau)$ of the $j$-th grain.
(Alternatively, one may think of $\phi_j$ as the time integral of
the {\it voltage} on the grains, $i \dot\phi_j= V_j$.)  The action
of the model contains two terms, $S = S_c+ S_{t}$, where
$S_c[\phi]= \sum_j \int\! d\tau [\dot\phi_j^2/(4E_c) -
iq\dot\phi_j]$, is the charging energy of a grain kept at voltage
$V_j = i\dot \phi_j$, and
\begin{equation}
  S_{t}[\phi] = {g T^2\over 2}\sum_{j=0}^{M-1}  \int\limits_0^\beta\!
  d\tau d\tau'\,
  {\sin^2(\Delta\phi_{j}(\tau)-\Delta\phi_{j}(\tau'))\over\sin^2(\pi
  T(\tau-\tau'))},
\label{aes}
\end{equation}
describes the process of tunneling. Here $\Delta \phi_j \equiv
(\phi_{j+1}-\phi_j)/2$ where $i \dot \phi_0$ and $i\dot \phi_{M}$ are
the voltages on the leads connected to the array.

Before analyzing the array in terms of the above action, let us review
a few general features of the AES approach: (i)  ignoring effects of
quantum interference, the applicability of the model is
restricted~\cite{Beloborodov01} to temperatures $T> g \delta$; (ii)
The quadratic approximation to the action, $S^{(2)}[\phi]= {1\over T}
\sum_{j,m} \left[{\omega_m^2\over 4 E_c }|\phi_{j}|^2 + 2g |\omega_m|
  |\Delta \phi_{j}|^2\right] $, provides a complete description of the
{\it classical} RC-resistor network corresponding to the array; (iii)
anharmonic fluctuations of the phase lead to the perturbative
logarithmic correction to the dc conductivity \cite{Golubev92}
mentioned in the introduction; (iv) technically, the field $\phi_j$
represents a mapping $S^{1}\to S^{1}$ from the unit circle (imaginary
time augmented with periodic boundary conditions) into itself
($\phi_j$ is a phase). In addition to $\phi=0$, the tunneling action
$S_t[\phi]$ of a single grain (which, for low temperatures $T\ll E_c$
represents a good approximation to the {\it total} action of the
grain) possesses a set of topologically non--trivial extremal phase
configurations known as Korshunov
instantons~\cite{Korshunov87,Zaikin91,Grabert96,Nazarov99}:
\begin{equation}
\exp(i\phi^{(z)}(\tau)) \equiv  \prod\limits_{\alpha = 1}^{|W|}
{e^{2\pi i \tau T}-z_\alpha \over 1- \bar z_\alpha e^{2\pi i \tau
T}}\,.
 \label{korshunov}
\end{equation}
Here, $W \in \Bbb{Z}\setminus 0$ is the winding number of the mapping
$\phi^{(z)}$ and $z\equiv (z_1,\dots,z_{|W|})$ is a set of $|W|$
complex parameters constrained by $|z_\alpha|<1$. The action
associated with the instanton, $S[\phi^{(z)}]\approx g|W|- 2\pi i q
W$, is nearly $z$--independent~\cite{foot2} which identifies the
$z_\alpha$'s as instanton zero modes. (Physically, $\arg{z_\alpha}$
determines the instance and $1-|z_j|$ the duration of the voltage
pulse, $i\dot\phi^{(z)}$.)

Turning to the array, the fact that the tunneling action depends
only on the {\it differences} of neighboring phases, $\Delta
\phi_j$, implies that a 'plateau' formed by $L$ instanton fields
embedded into $M-L$ zeros,
$(0,\dots,0,\phi^{(z)},\dots,\phi^{(z)},0,\dots,0)$, represents an
extremal configuration. For $W=\pm 1$ its action is given by
$S[\phi]=L (\pi^2 T/E_c \mp 2\pi i q) + g\,$, an expression that
suggests an alternative interpretation of the instanton plateau:
rather than monitoring a state of every grain, one may think of
the plateau as a dipole of two charges located  at the positions
of the step--wise changes in the winding number, $W_j$: $0\to 1$
and $1\to 0$, resp. Within this picture, $\exp(-g/2)$ represents
the fugacity of the charges, $|L|\pi^2T/E_c$, their interaction,
and the $q$-dependent term describes the interaction of the dipole
with a uniform electric field $2\pi i q$. More formally, a
summation over all instanton configurations followed by
integration over massive Gaussian fluctuations and zero
modes~\cite{foot5,Altland} leads to the expression
\begin{equation}
\frac{\cal Z}{{\cal Z}_0} =\sum\limits_{k=0}^{\infty} \frac{\left({\gamma E_c\over T}\right)^{2k}}{(k!)^2} \!
\sum\limits_{j_1\ldots j_{2k}}^{M-1} e^{-{1\over 2}\sum\limits_{a,b}^{2k} V(j_a-j_{b}) - \sum\limits_a^{2k} \Phi(j_a) }  ,
                                                     \label{partition}
\end{equation}
where $\gamma^2\equiv g^{3} e^{-g}$, and the interaction
potentials are
\begin{equation}
V(j_a-j_b) =  {\pi^2 T\over E_c}\, e_ae_b |j_a-j_b| \,; \,\,\,\,
\Phi(j_a) = 2\pi i qe_a j_a \, ,
                                                      \label{potential}
\end{equation}
with $e_a \equiv (-1)^a$. These equations generalize from a single
dipole to the statistical mechanics of a 1$d$ Coulomb gas in a uniform
external field. The fugacity of the gas, $\gamma E_c/T$, results from
multiplication of the instanton action by the fluctuation
factor~\cite{foot3}.

To understand the properties of this system, we recall the
standard mapping of a Coulomb gas onto the sine-Gordon
model~\cite{map}.  In the present context, the action of the
latter is given by Eq.~(\ref{eq:2}), which completes the proof of
equivalence of the two approaches discussed in this Letter upon
the proper identification of $\gamma$~\cite{foot4}. Therefore the
activation temperature, $T^*(q)$, is given by Eq.~(\ref{tstar})
with $\gamma=g^{3/2}e^{-g/2}$. It is worthwhile to mention that a
key element in establishing this equivalence is the factor $E_c/T
\gg 1$ in the fugacity of the Coulomb gas, Eq.~(\ref{partition}).
This factor results from the large volume available to
fluctuations in the array~\cite{foot3} geometry, i.e. no such
factor exists for single grains.

We finally turn to the discussion of the low temperature
($T<T^\ast$) dc transport properties of the array. As mentioned
above, in the insulating phase the fundamental excitations of our
system are solitary configurations carrying unit charge. Referring
for detailed discussion to Ref.~\cite{Altland}, we here merely
mention that in the presence of an external field, $E$, the
dynamics of these objects is controlled by the Langevin equation
\begin{equation}
{\partial \theta_j \over \partial t}- g E_c \left[ {\partial^2
\theta_j \over
\partial j^2} - \gamma \sin(\theta_j + jq)\right] = gE + \xi(t)\, ,
                                     \label{Langevin}
\end{equation}
where $\xi(t)$ is a Gaussian correlated noise with $ \langle \xi(t)
\xi(t') \rangle_\xi = gT\, \delta(t-t')$ and ${\partial^2
  \theta_j/\partial j^2}\equiv \theta_{j+1}-2\theta_j+\theta_{j-1}$ is
the discrete second derivative.

In the commensurate phase ($|q| <q^\ast$) the solutions of this
equation are solitary configurations, $\theta_j(t) =
\tilde\theta(j-vt)$, propagating with a constant velocity, $v$.
Substituting this ansatz into Eq.~(\ref{Langevin}), one finds $ v
=\gamma^{-1/2}gE$, where $\gamma^{-1/2}\gg 1$ is the soliton
length. As each of these objects carries unit charge, the current
density is given by $J=en v$, where $n= \gamma^{1/2}
e^{-T^\ast/T}$  is the concentration of the thermally excited
solitons ($T^\ast$ is given by Eq.~(\ref{tstar}) with
$\gamma=g^{3/2}e^{-g/2}$). The linear dc conductivity of the array
is thus given by Eq.~(\ref{result}). In the case of a  random
background charge, $q_j$, we expect a similar result with,
however, a different activation energy $T^\ast\sim
E_c\gamma\approx E_c g^{3/2}e^{-g/2}$.
The linear $I$--$V$ characteristics breaks down once the voltage drop
per grain exceeds some critical value; even in the case of the
largest energy gap ($q=0$), this value is fairly low, $V_c \approx
\gamma E_c\ll E_c$.

Summarizing, we have considered a 1$d$ array of metallic grains
connected by highly conducting junctions.  We have shown that the
inelastic tunneling and weak charge quantization lead  to
insulating behavior below temperatures $T^\ast\sim E_c e^{-g/4}$
(array with no background charges) or $T^\ast\sim E_c e^{-g/2}$
(random array).  Both scales are much larger than the energy
$\tilde{E}_c\sim E_c e^{-g}$, where perturbative mechanisms
inhibiting charge transport become sizeable. In essence this
phenomenon is explained by the  analogy between the array and an
elastic "chain" pinned by a periodic potential. Most importantly,
even an exponentially weak pinning potential leads to the
formation of a "commensurate" phase where  the thermodynamic DOS
and the linear conductivity exhibit activation behavior.  The
mechanism discussed here may play a important role in the
construction of an "extended" theory encompassing both strong
interaction and effects of long range quantum interference.

We are indebted to A. I. Larkin and J. Meyer for valuable
discussions. This work is supported by NSF grants DMR97-31756,
DMR01-20702, DMR02-37296, and EIA02-10736.


\begin{references}
\bibitem{foot2}
  This is true for $1-|z_\alpha|^2\gg (T/E_c)$, which is always the
  case in our calculations.
\bibitem{foot5} Doing so, we neglect temporal correlations (interactions) between
  instantons with $|W|>1$ on a given grain. This approximation is
justified for $T\gg \tilde E_c$.
\bibitem{foot3} The massive Gaussian fluctuations bring the
  factor $(gE_c/T)^{2k(1-1/M)}$ \cite{Altland}, which for the single dot
  ($M=2$, $k=1$) reads as $g E_c/T \ln(E_c/T)$ \cite{Grabert96}.  An
  additional factor of $g^k$ originates from the zero--modes integrations.
\bibitem{foot4} If transparency of all channels is small
  $|t_s|^2=1-|r_s|^2\ll 1$, one finds $\gamma\sim \exp\{{1\over
    2}\sum_s \ln(1-|t_s|^2)\}\approx \exp\{-g/2\}$, where
  $g\equiv \sum_s|t_s|^2$, i.e.  the two definitions of $\gamma$ agree with
  exponential accuracy.
\end{references}
\end{document}